\documentstyle[prd,multicol,aps,tighten]{revtex}
\def\bea{\begin{eqnarray}}
\def\eea{\end{eqnarray}}
\def\be{\begin{equation}}
\def\ee{\end{equation}}
\begin{document}


\title{{\bf Holographic Stress Tensors for Kerr--AdS Black Holes}}

\author{Adel M. Awad{$^\sharp$} 
and Clifford V. Johnson$^{\sharp,\flat}$}

\address{\hfil}
\address{\cite{cvjone}Department of Physics and Astronomy,
University of Kentucky,  Lexington, KY 40506, U.S.A.}
\address{\cite{cvjtwo}School of Natural Sciences, Institute for 
Advanced Study, 
Olden Lane, Princeton, NJ 08450, U.S.A.}
\address{and}
\address{Centre for Particle Theory, Department of Mathematical Sciences,
University of Durham, Durham, DH1 3LE, U.K.}
\address{\hfil\\\tt adel@pa.uky.edu, c.v.johnson@durham.ac.uk}

\address{\hfil}

\date{September 1999}
\maketitle

\begin{abstract} We use the counterterm subtraction method to
calculate the action and stress--energy--momentum tensor for the (Kerr)
rotating black holes in AdS$_{n+1}$, for $n${=}2,3~and~4. We
demonstrate that the expressions for the total energy for the
Kerr--AdS$_{3}$ and Kerr--AdS$_{5}$ spacetimes, in the limit of
vanishing black hole mass, are equal to the Casimir energies of the
holographically dual $n$--dimensional conformal field theories. In
particular, for Kerr--AdS$_5$, dual to the case of four dimensional
$N{=}4$ supersymmetric Yang--Mills theory on the rotating Einstein
universe, we explicitly verify the equality of the zero mass stress
tensor from the two sides of the correspondence, and present the
result for general mass as a prediction from gravity.  Amusingly, it is
observed in four dimensions that while the trace of the stress tensor
defined using the standard counterterms does not vanish, its integral
does, thereby keeping the action free of ultraviolet
divergences. Using a different regularisation scheme another
stress tensor can be defined, which is traceless.

\end{abstract}

\begin{multicols}{2}

\section{Introduction} The AdS/CFT correspondence relates an
$(n{+}1)$--dimensional theory of gravity on anti--de Sitter (AdS)
spacetime (times a compact manifold) to a conformal field theory (CFT)
in $n$ dimensions. This duality first arose as a result of
investigating\cite{Maldacena} $N$ parallel D3--branes in the context
of the low energy, ({\it i.e.,} the limit of zero~$\alpha^\prime$, the
inverse string tension) classical (weak string coupling,~$g_s$) limit
of type~IIB superstring theory on AdS$_5{\times}S^5$. A precise
statement of the AdS/CFT correspondence\cite{Witten,Gubser} is the
equality of the partition functions of the two theories,
\bea
Z_{AdS}(\phi_{i})=Z_{CFT}(\phi_{0,i})\ .
\eea
{}From the gravity--on--AdS point of view, $\phi_{i}$ is a bulk field
constrained to the values $\phi_{0,i}$ on the boundary, while from the
CFT point of view, $\phi_{0,i}$ are sources for pointlike operators,
${\cal O}_i$, in the theory. In the low energy limit of the theory one
can use the classical gravitational action to calculate the partition
function of the CFT on the boundary. This action has the
form\cite{Gibbons},
\bea
I_{\rm bulk}+I_{\rm surf}=
&-&{1 \over 16 \pi G}\int_{\cal M} d^{n+1}x
\sqrt{-g}\left(R+{n(n-1) \over l^2}\right)\nonumber\\
      &-&{1 \over 8 \pi G } \int_{\partial {\cal M}} d^{n}x \sqrt{-h} \,K.
\eea
The first term is the Einstein--Hilbert action with negative
cosmological constant ($\Lambda{=}{-n(n{-}1)/2l^2}$). The second term
is the Gibbons--Hawking boundary term.  Here, $h_{ab}$ is the boundary
metric and $K$ is the trace of the extrinsic curvature $K^{ab}$ of the
boundary.

To deal with the divergences which appear in the gravitational action
(arising from integrating over the infinite volume of spacetime), two
different techniques may be employed. The ``traditional'' background
subtraction technique\cite{Gibbons,Hawkingone} (which subtracts the
contribution from a reference spacetime to get a finite result) and
the ``counterterm subtraction'' method\cite{Balasubramanian} (which
regulates the action by the addition of certain boundary counterterms).
As the counterterms depend upon the geometrical properties of the
boundary of the spacetime, the counterterm subtraction method provides
an intrinsic definition of the action for a particular spacetime.
This sidesteps problems which arise in using the other method when the
spacetime in question has an ambiguous (or simply unknown) choice of
background.

The divergences of the Einstein--Hilbert action (in dimensions less
than six) can be cancelled by adding the following
counterterms:
\bea
I_{\rm ct}={1 \over 8 \pi G}
 \int_{\partial {\cal M}}d^{n}x\sqrt{-h}\left[ \frac{(n-1)}{ l}-{l{\cal R}
\over 2(n-1)}\right].
\label{theterms}
\eea
Here ${\cal R}$ is the Ricci scalar for the boundary metric $h$.
Using these counterterms one can construct a divergence--free stress tensor
which is given
by ($I{=}I_{\rm bulk}{+}I_{\rm surf}{+}I_{\rm ct}$):
\bea
T^{ab}&=& {2 \over \sqrt{-h}} {\delta I \over \delta h_{ab}}=\nonumber\\
&&{1 \over 8 \pi G}\left[K^{ab}-h^{ab} K-\frac{(n-1)}{l} h^{ab}
     +{lG^{ab} \over (n-2)}\right]\ ,
\label{stressone}
\eea
where $G^{ab}$ is the Einstein tensor in $n$
dimensions. (Correspondingly, the last term should be omitted for
$n{=}2$.)

(In ref.\cite{counter}, the counterterm subtraction method was studied
extensively, with many examples, and a new counterterm was presented
there which allows the subtraction regularisation to be performed in
dimensions $n{+}1{=}6,7$. See also
refs.\cite{lau,mann,charged,roberto,kraus,klemm,solo} for related
studies, some more applications and further extensions.)

The prescription (\ref{stressone}) gives a definition of the action
and stress--tensor on any region (say, of radius $r$ in the
coordinates that we will choose later) bounding the interior of
AdS. The AdS/CFT holographic relation (in the form which we will need
it here) equates these quantities to a dual conformal field theory
residing on the boundary at infinity ($r{\to}\infty$).  The theory on
the boundary at radius~$r$ can be taken to be a dual field theory with
an ultraviolet cutoff proportional to~$r$. Then $r{\to}\infty$ defines
the UV fixed point conformal field theory. This dovetails nicely with
the fact that the counterterms, while regulating an infra--red (IR)
divergence coming from the bulk, have the dual interpretation as
regulating UV divergences in the field
theory\cite{wittensussk,Balasubramanian}.

Recalling that the metric restricted to the boundary, $h_{ab}$,
diverges due to an infinite conformal factor $r^2/l^2$, we define the
background metric upon which the dual field theory resides as
\be\gamma_{ab}=\lim_{r\to\infty}{l^2\over r^2}h_{ab}\ .
\label{newmetric}
\ee
Consequently, the field theory's stress--tensor, ${\widehat T}^{ab}$,
is related to the one above (\ref{stressone})
 by the rescaling\cite{robstress}:
\be
\sqrt{-\gamma}\,\gamma_{ab}{\widehat
T}^{bc}=\lim_{r\to\infty}\sqrt{-h}\,h_{ab}T^{bc}\ ,
\label{newstress}
\ee which amounts to multiplying all expressions for $T^{ab}$
displayed later by $(r/l)^{n-2}$ before taking the limit
$r{\to}\infty$.

In this paper we apply these techniques to the study of the
Kerr--AdS$_{n+1}$ spacetimes\footnote{See
refs.\cite{Hawkingtwo,Berman,harvey} for complementary studies of
higher dimensional Kerr--AdS spacetimes and their field theory
duals.}, for $n{=}2,3$ and $4$. In particular, we compute the action,
stress tensor, and other quantities, and consider their implications
for the dual field theories. The dual field theories are defined on
the rotating spacetimes located at the boundary of Kerr--AdS, and the
results which we present are new for the three and four dimensional
cases.

In fact, for the massless case we exactly reproduce the four
dimensional result for the stress tensor using field theory.  The
results for general mass are offered as predictions from the gravity
side about the strongly coupled field theory at finite
temperature. Along the way, we learn a number of interesting things
about the dual field theories in three and four dimensions. In
particular, the trace of the four dimensional stress tensor obtained
using the counterterms~(\ref{theterms}) ---for ${\cal N}{=}4$
supersymmetric Yang--Mills on the rotating Einstein universe--- does
not vanish, but nevertheless integrates to zero.  However, we can
define a new stress tensor which is traceless(however, see note added). This
illustrates the
ambiguity of the definition of the trace anomaly in four dimensions,
allowing a term proportional to $\Box{\cal R}$ to be added. In the
non--rotating case, such a term is not needed\cite{Balasubramanian},
and so we find here that with rotation present, a different
regularisation scheme must be used~\cite{Birrel}, corresponding to
supplementing the counterterms (\ref{theterms}) with a term
proportional to $\int_{\partial{\cal M}}{\cal R}^2$. (This possibility
was noted in ref.\cite{Balasubramanian}, and we believe that this is
the first such example in this AdS/CFT context.)

\section{Kerr--AdS$_{3}$} As a warm--up and review, we will study the
case of Kerr--AdS$_3$. The explicit computation of the
energy--momentum tensor and the mass, angular momentum and Casimir
energy using the counterterm subtraction technique has been done in
ref.\cite{Balasubramanian}. Here, we will perform all of those
computations in a different choice of coordinates, supplementing the
discussion and calculations where necessary.  This will serve the twin
purposes of setting up the notation of the rest of the paper, and
illustrating the similarities to (and differences from) the higher
dimensional cases which we later present.

We use the form of Kerr--AdS$_{3}$ metric in ref. \cite{Hawkingtwo}
which resembles the higher dimensional Kerr--AdS metrics and can be
obtained from the BTZ black hole by coordinate
transformation\cite{btz,bthz}:
\bea
ds^2&=&-{\Delta_{r} \over r^2}\left(dt-{a \over \Xi } d\phi\right)^2 +{r^2
\over
\Delta_{r}}dr^2\nonumber\\
    & &+{1 \over r^2}\left(adt-{(r^2+a^2) \over \Xi } d\phi\right)^2\ ,
\label{adsthree}
\eea
where
\bea
\Delta_{r}&=&(r^2+a^2)\left(1+\frac{r^2}{l^2}\right)-2MGr^2,\\
       \Xi&=&\left(1-\frac{a^2}{l^2}\right)\ .
\eea
Here, $a$ is the rotational parameter. The horizons are the zeros of
${\Delta}_{r}$, and
$r_{+}$ is the outer horizon which is\cite{bthz}:
\bea
r_{+}^2&=&{l^2 \over 2}\left(2MG-1-\frac{a^2}{l^2}\right)\nonumber\\
     & &+{l^2\over 2}
\sqrt{\left(1+\frac{a^2}{l^2}-2MG\right)^2-4\frac{a^2}{l^2}}\ .
\eea

If we were to take $t{\rightarrow}i\tau$, defining the Euclidean
section (putting the theory at finite temperature), regularity
(thermal equilibrium) requires that the period of Euclidean time
($\tau{=}it$) ---the inverse temperature $\beta$--- has the form: \bea
\beta={2 \pi (r_{+}^2+a^2)r_{+} \over (r_{+}^{4}/l^2-a^2)}\ .  \eea
The angular velocity of the horizon is \bea \Omega={a \Xi \over
r_{+}^2+a^2}\ ,
\label{angular}
\eea while the area of the horizon is \be {\cal A}={2 \pi
(r_{+}^2+a^2) \over r_{+}\Xi}\ .  \ee In the Euclidean continuation,
where we must also Wick rotate $a$ to $i\alpha$, as the horizon is a
bolt of the Killing vector $\partial_\tau+i\Omega\partial_\phi$, the
identification $\tau{\sim}\tau{+}\beta$ results\cite{Hawkingtwo} in
the identification $\phi{\sim}\phi{+}i\beta \Omega$, in addition to
the usual $\phi{\sim}\phi{+}2\pi$.

The boundary metric of Kerr--AdS$_{3}$ is, as $r{\to}\infty$, given by
\be ds^2= \frac{r^2}{l^2}\left[-dt^2+{2a \over \Xi}dtd\phi +{l^2 \over
\Xi}d\phi^2\right]\ .  \ee Removing the conformal factor, our dual
field theory is defined on the spacetime with metric $\gamma_{ab}$:
\be [\gamma_{ab}]=\pmatrix{-1&{a/\Xi}\cr{a/\Xi}&{l^2/\Xi}}\ .  \ee

After some computation, the components of the stress tensor at large $r$ are
found to be
\bea
 8\pi GT_{tt}&=&{1 \over 2} \left(2MG-2+\Xi\right)+O\left({1 \over
r}\right),\nonumber\\
 8\pi GT_{t\phi}&=&-{a(2MG+\Xi) \over 2l\Xi)}+O\left({1 \over
r}\right),\nonumber\\
  8\pi GT_{\phi\phi}&=&{(2MG(1+a^2/l^2)-\Xi^2)l \over
2\Xi^2} +O\left({1 \over r}\right).
\label{formone}
\eea

It is interesting to note that the resulting field theory stress
tensor (obtained using eqn.(\ref{newstress}) and discussion below) can
be written in the following form\footnote{We thank R.~C.~Myers for the
suggestion that ${\widehat T}^{ab}$ might be written in this form.}:
\be {\widehat T}^{ab}=A\left(2u^a u^b+\gamma^{ab}\right) +B v^a v^b
\label{formtwo} \ee where $ u^a{=}(1,0)$ is a unit time--like two
vector and $v^a{=}(1, 1/l{-}a/l^2)$ is a null vector, {\it i.e.,}
$v^bv_b{=}0$. The tensor is therefore manifestly traceless.  The
coefficients are: \bea A = { \left(2MG-(1+{a/l})^2 \right) \over 16\pi
Gl};\quad B = {a(1+a/l)^2\over 8\pi Gl^2}\ .  \eea It is
straightforward to check that this tensor is covariantly conserved.
Notice that when $a{=}0$, $B$ vanishes and $A{=}(2MG{-}1)/(16\pi Gl)$,
giving a standard form for the stress--energy--momentum tensor of a
fluid of massless particles with some energy density given by
${\widehat T}_{00}$.  Notice that $A$, in front of the part of the
stress tensor in standard form contains the black hole parameter $M$
---the thermal part of the field theory--- while $B$, the coefficient
of the null vector part, refers only to rotational parameters. We will
find that this form will persist to higher dimensions.

To calculate the conserved quantities for these spacetimes, we use the
following definition for a conserved charge \cite{Brown}, associated
to a symmetry generated by the Killing vector $\xi^\mu$:
\be
Q_{\xi}=\int_{\Sigma}d^{n-1}x\sqrt{\sigma} u^{\mu}T_{\mu\nu}\xi^{\nu}\ .
\label{charges}
\ee
where $u_{\mu}{=}-N\,t,_{\mu}$, while $N$ and $\sigma$ are the lapse function
and the
spacelike metric which appear in the ADM--like decomposition of the boundary
metric
\be
ds^2=-N^2 dt^2+\sigma_{ab}(dx^{a}+N^a dt)(dx^b+N^b dt)\ .
\ee
Our convention for the Killing vectors $\xi^\mu$ is as follows:
$\partial_t$ is the Killing vector conjugate to the time $t$ and
$\partial_{\phi}$ is the Killing vector conjugate to $\phi$. Using the
above definition for the conserved charge, the mass and the angular
momentum of the Kerr--AdS$_{3}$ spacetime are given by
\be
{\cal M}={1 \over 8G\Xi}\left(2MG-\Xi\right)\ ,\quad
{\cal J}={1 \over 2}{M a \over \Xi^2}\ .
\ee
Direct evaluation of the counterterms gives the finite action
\bea
I_{3}=I_{\rm bulk}+I_{\rm surf}+I_{\rm ct}=-{1 \over 4G}{\pi(r_{+}^2+a^2)
 \over
r_{+}\Xi}\ .
\eea
The action with the other  quantities satisfy the following thermodynamical
relation
\be
S=\beta ({\cal M}-\Omega {\cal J})-I_{n+1}={{\cal A} \over 4 G}\ ,
\label{firstlaw}
\ee
for $n{=}2$, which is a non--trivial check of some of our
computations. We will perform this check in the more complicated
examples to come.

\subsection{Comparison to Field Theory}
Our dual field theory resides on the spacetime with metric $\gamma_{\mu\nu}$
giving line element
\bea
ds^2= -dt^2+{2a \over \Xi}dtd\phi +{l^2 \over \Xi}d\phi^2\ .
\eea
Notice that this can be brought into the form of a metric on a cylinder
$(R{\times}S^{1})$
\be
ds^2=-dT^2+ R^2\,d\Phi^2\ ,
\ee
with $R{=}l/\sqrt{\Xi}$,
using the following coordinate transformation\cite{Hawkingtwo}:
\be
T= {t \over \sqrt{\Xi}},\,\, \Phi= \phi+at/l^2\ .
\label{coords}
\ee

Our traceless result above (eqn.'s (\ref{formone}) and (\ref{formtwo}))
therefore correctly reproduces the result for the trace anomaly, given
in two dimensions by
\bea
{\widehat T}^{a}_{a}=-{c {\cal R } \over 24\pi}\ ,
\eea
where $c$ is the central charge and ${\cal R}$ is the intrinsic curvature of
spacetime,
which is zero for the cylinder.

According to the correspondence, the Casimir energy of the dual field
theory\footnote{In higher dimensions, connections between energy of a
spacetime solution and Casimir energy of a dual field theory were
first pointed out in ref.\cite{robgary}.} is the contribution to the mass of
spacetime $\cal M$ which is independent of the black hole
parameter, $M$, and is given
by \be {\cal E}=-{1 \over 8G}\ .
\label{cassy}
\ee Given the standard result~\cite{henneaux} for the relationship
between the central charge of the conformal field theory and gravity
in AdS$_3$, $c{=}{3 l/ 2G}$, we see that this translates into a vacuum
energy $-c/(12l)$ for the theory on the cylinder. This is consistent
with the interpretation~\cite{juanandy} of the
spacetime~(\ref{adsthree}), with $M{=}0$, as the
Neveu-Schwarz--Neveu-Schwarz vacuum of the holographically dual super
conformal field theory: The fermions have anti--periodic
boundary conditions as they go once around the cylinder, preventing
their zero--point energy from cancelling that of the bosons, as 
happens in the Ramond--Ramond sector.

\section{Kerr--AdS$_{4}$}

The Kerr--AdS$_{4}$ metric has the following form\cite{carter}
\bea
ds^2&=&-{\Delta_{r} \over \rho^2}\left(dt-{a \sin^2{\theta} \over \Xi }
d\phi\right)^2
+{\rho^2 \over \Delta_{r}}dr^2+{\rho^2 \over\Delta_\theta}d\theta^2\nonumber\\
    & &+{\Delta_{\theta} \sin^2{\theta} \over \rho^2}\left(adt-{(r^2+a^2)
\over
\Xi} d
\phi\right)^2\ ,
\label{adsfour}
\eea
where
\bea
\Delta_{r}&=&(r^2+a^2)(1+r^2/l^2)-2MrG\ ,\nonumber\\
\Delta_{\theta}&=&1-(a^2/l^2)\cos^2{\theta}\ ,\nonumber\\
\rho^2&=&r^2+a^2\cos^2{\theta}\ .
\label{metricstuff}
\eea The period of Euclidean time is given by \bea \beta= {4 \pi
(r_{+}^2+a^2) \over r_{+}(3r_{+}^2/l^2+1+a^2/l^2-{a^2/ r_{+}^2})}\ ,
\eea with the angular velocity of the horizon 
given by eqn.~(\ref{angular}).  
Here, $r_+$ is the location of the horizon,
the largest root of $\Delta_r$. 
Again, after continuing $a{\to}i\alpha$, Euclidean regularity gives
$\phi$ a period $\phi{\sim}\phi{+}i\beta \Omega$, in addition to the
usual $\phi{\sim}\phi{+}2\pi$.

The area of the horizon is
\bea
{\cal A}=4 \pi \left({r_{+}^2+a^2 \over \Xi}\right)\ .
\eea

The non--vanishing components of the Kerr--AdS$_4$ stress tensor for
large $r$ exactly match the components recently computed for the
Kerr--Newman--AdS$_4$ case in ref.\cite{klemm} in the limit where the
charges vanish ({\it i.e.,} their~$z{\rightarrow}0$) and are
\bea
8\pi GT_{tt}&=&{ 2M \over r l}+O\left({1\over r^2}\right)\\
       8\pi GT_{t \phi}&=&-{2a M \over r \Xi l}\sin^2{\theta}+O\left({1\over
r^2}\right)\\
 8\pi GT_{\theta\theta}&=&{M l\over r \Delta_{\theta}}+ O\left({1\over
r^2}\right)\\
    8\pi GT_{\phi\phi}&=&{Ml\sin^2{\theta} \over r
\Xi^2 }[3a^2\sin^2{\theta}/l^2+\Xi]+O\left({1\over r^2}\right).
\eea
The mass and the angular momentum  are computed from this as:
\bea
{\cal M}={M \over \Xi };\,\,\,
{\cal J}={a M \over \Xi^2}\ .
\eea

The boundary metric of Kerr--AdS$_{4}$ is given by 
\be
ds^2=\frac{r^2}{l^2}\left[-dt^2+{2a\sin^2{\theta}\over\Xi}dtd\phi
+l^2{d\theta^2
\over \Delta_{\theta}}+l^2{\sin^2{\theta} \over \Xi}d\phi^2\right]\ .
\ee

Removing the conformal factor (see (\ref{newmetric})), our dual field
theory is defined on the spacetime (with coordinates
$(t,\phi,\theta)$) with metric $\gamma_{ab}$: \be
\left[\gamma_{ab}\right]=
\pmatrix{-1&{a\sin^2\theta/\Xi}&0\cr
{a\sin^2\theta/\Xi}&{l^2\sin^2\theta/\Xi}&0\cr0&0&l^2/\Delta_\theta}.
\ee

Converting with the conformal factor and taking the limit $r \to \infty$ (see
(\ref{newstress})), we find that the stress tensor of the field theory
has this simple form: 
\be 
{\widehat T}^{ab}={M\over 8\pi l^2}\left[3u^a u^b+\gamma^{ab}\right] 
\ee 
where $u^a{=}(1,0,0)$. This is
the standard form for the stress tensor of the non--rotating theory
2+1 dimensional conformal field theory!  The tensor is covariantly
conserved and manifestly traceless, the latter result being consistent
with the absence of a conformal anomaly in odd dimensional spacetime.
It is interesting to note that the tensor can be written in such a
simple form for this theory, even in the presence of
 rotation, in contrast to the
case in two dimensions (see eqn.(\ref{formtwo})), and as we will see,
the four dimensional case.

The action calculation in this case gives the result:
\be
I_{4}=-{\pi (r_{+}^2+a^2)(r_+^2/l^2-1) \over
G(3r_+^4/l^2+r_+^2+a^2r_{+}^2/l^2-a^2)\Xi}\ ,
\ee
which agrees with action calculation using the background subtraction
technique in ref.\cite{Hawkingtwo}(see also ref.\cite{mann}), since the
Casimir energy is zero for an odd dimensional field theory. Also it agrees
with the action for Kerr--Newmann~\cite{klemm}\ in the limit where the
charges vanish.  These quantities also satisfy the first law ({\it
i.e.,} eqn.(\ref{firstlaw}) with $n{=}3$.)

\section{Kerr--AdS$_{5}$}

The metric for the Kerr--AdS$_{5}$  in general has two rotation parameters
since the rotation group is $SO(4) {\cong} SU(2)_L {\times} SU(2)_R$. Here we
discuss the
one--parameter  solution  given by~\cite{Hawkingtwo}
\bea
ds^2&=&-{\Delta_{r} \over \rho^2}\left(dt-{a \sin^2{\theta} \over \Xi }
d\phi\right)^2 +r^2
\cos^2{\theta} d\psi^2+{\rho^2\over\Delta_\theta}d\theta^2\nonumber\\
    & &+{\rho^2 \over
\Delta_{r}}dr^2+{\Delta_{\theta}\sin^2{\theta}\over\rho^2}\left(adt-{(r^2+a^2)
\over \Xi}
d\phi\right)^2.
\label{adsfive}
\eea
Now we have
\be
\Delta_{r}=(r^2+a^2)(1+r^2/l^2)-2MG\ ,
\ee
and the remaining quantities are as in eqn.(\ref{metricstuff}).
This time the inverse temperature is
\be
\beta={2 \pi (r_{+}^2+a^2) \over r_{+}(2r_{+}^2/l^2+1+a^2/l^2)}\ ,
\ee
with the angular velocity of the horizon 
given again by eqn.~(\ref{angular}).  Again,
$r_+$ is the location of the horizon, the largest root of $\Delta_r$, and
$\phi$ has period $\phi{\sim}\phi{+}i\beta \Omega$, in addition to the
usual $\phi{\sim}\phi{+}2\pi$.

In ref.\cite{Hawkingtwo} the calculation for the relevant physical
quantities was carried out using the subtraction technique. The
reference spacetime in those calculations was the spacetime with
$M{=}0$, {\it i.e.,} AdS$_5$ in very non--standard coordinates.  Here, we
go some steps further, by computing and studying physics intrinsic to
the Kerr--AdS$_5$ spacetime with no reference to a background, 
allowing us to extract physical quantities like the Casimir
energy and other interesting features of the stress tensor, as we shall see.

After computation, we get the following non--vanishing components for the
stress tensor at large $r$,
\bea
8\pi G T_{tt}&=&{l \over 8r^2}[24MG/l^2-14a^{2}\cos^2\theta
/l^2-14a^{4}\cos^2\theta/l^4\nonumber\\
           & &+15a^{4}\cos^4\theta/l^4+3(1+a^2/l^2)^2]
                   +O\left({1\over r^4}\right)\ ,
\nonumber\\
8\pi G T_{t\phi}&=&{al\sin^2\theta \over 8r^2 
\Xi}[2a^2\cos^2\theta/l^2-7a^4\cos^4\theta/l^4+\Xi^2\nonumber\\
                   & &+10a^4\cos^2\theta/l^4-24MG/l^2-4a^4/l^4]\nonumber\\&&
\hskip5cm + O\left({1\over r^4}\right)\ ,\nonumber\\
8\pi G T_{\phi\phi}&=&{l^3\sin^2\theta \over \Xi^2}[7a^6\cos^4/l^6
-7a^4\cos^4\theta /l^4+3a^6/l^6\nonumber\\ & &-8a^4\cos^2\theta/l^4
  -32a^2MG\cos^2\theta/l^4+1\nonumber\\
& &-10a^6\cos^2\theta\l^6-3a^2/l^2+2a^2\cos^2\theta/l^2\nonumber\\
&&+24a^2MG/l^4-a^4/l^4+8MG/l^2]+ O\left({1\over
r^4}\right) ,\nonumber \\
8\pi G T_{\theta\theta}&=&{l^3 \over 8 r^2
\Delta_{\theta}}[8MG/l^2+\Xi^2-3a^4\cos^4{\theta}/l^4\nonumber\\
                      &
&+2a^2\cos^2{\theta}/l^2+2a^4\cos^2{\theta}/l^4]+ O\left({1\over
r^4}\right)\ ,
\nonumber\\
8\pi G T_{\psi\psi}&=&{l^3\cos^2{\theta} \over 8
r^2}[2a^{2}\cos^2{\theta}/l^2+\Xi^2+8MG/l^2\nonumber\\
                 &
&-7a^{4}\cos^4{\theta}/l^4+2a^{4}\cos^2{\theta}/l^4
            ]+O\left({1\over
r^4}\right).
\eea
Using the above definition (\ref{charges}) for a conserved charge, one
can calculate the mass and angular momentum of the solution, with the
result:
\bea
{\cal M}={\pi l^2\over 96 G \Xi}[a^4 /l^4+9 \Xi+72 G M/l^2 ]\ ,\quad
{\cal J}={\pi M a \over 2 \Xi^2}\ .
\label{energy}
\eea
For the action one gets
\bea
I_{5}&=&{\pi^{2}l^2(r_{+}^{2}+a^{2}) \over 48G
r_{+}(2r_{+}^{2}/l^2+1+a^{2}/l^{2})\Xi}[a^{4}/l^{4}
-12(r_{+}^{2}a^{2}/l^4\nonumber\\
     & &+r_{+}^{4}/l^4-r_{+}^{2}/l^2)+9+3a^{2}/l^{2}]\ .
\label{theaction}
\eea
The area of the horizon is
\bea
{\cal A}=2 \pi^2 {r_{+}(r_{+}^2+a^2) \over \Xi}\ .
\eea
The above quantities satisfy the first law ({\it i.e.,} eqn.(\ref{firstlaw})
with $n{=}4$).

\subsection{Comparison to Field Theory}
The metric on the boundary is that of a rotating Einstein
universe\cite{cassidy}: \bea ds^2&=& \frac{r^2}{l^2}
\left[-dt^2+{2a\sin^2{\theta}\over\Xi}dtd\phi+l^2{d\theta^2 \over
\Delta_{\theta}}\right.\nonumber\\ & &
\hskip2cm\left.+l^2{\sin^2{\theta} \over \Xi}d\phi^2+l^2\cos^2{\theta}
d \psi^2\right]\ .
\label{rotatestein}
\eea

Removing the factor $r^2/l^2$ (defining the metric $\gamma_{ab}$)
gives the line element of the space--time (with coordinates
$(t,\phi,\theta,\psi)$) upon which our conformal field theory resides,
and for which we must compute the energy--momentum tensor in order to
compare to the gravity computation. This seems at first a daunting
prospect, until one notices by direct computation that the Weyl tensor
vanishes for this spacetime, showing that it is conformally flat. This
indeed follows from the fact\cite{Hawkingtwo} that the spacetime
(\ref{adsfive}), with $M{=0}$, is actually just AdS$_5$ in
non--standard coordinates, and so its boundary shares some of the
conformal properties of the boundary of AdS$_5$.

One can therefore use the following general expression from
ref.~\cite{Birrel} to calculate 
 the stress tensor of a field theory defined on
conformally flat spacetime in four dimensions:
\bea
<\!{{\widehat T}^{s}}_{ab}\!>=-{1\over 16 \pi^2}\left[{1\over 9} \alpha^{s}
{H_{ab}}^{(1)}+ 2
\beta^s {H_{ab}}^{(3)}\right]\ ,
\label{stresstwo}
\eea
where ${H_a^b}^{(1)}$, ${H_a^b}^{(3)}$, $\alpha^s$ and $\beta^s$ are
defined in ref.\cite{Birrel} (There, they use spacetime indices
$(\mu,\nu)$ for the field theory while here we use $(a,b)$).  The
label $s{\in}\{0,1/2,1\}$ distinguishes the spin of the field for
which the labelled coefficients $\alpha^s$ and $\beta^s$ are
computed. We can now compare this to the the non--thermal ({\it i.e.,}
the $M$--independent) part of the tensor which we have computed on the
gravity side.

According to the AdS/CFT holographic relation, one should define the
Casimir energy for the field theory dual to the Kerr--AdS spacetime as
the contribution to the total energy of the spacetime (\ref{energy}) which
is independent
of the black hole's mass. This is given by
\be
{\cal E}={\pi l^2 (a^4/l^4+9\Xi) \over 96\, G \Xi}\ .
\label{casimirtwo}
\ee
In the limit $a{\rightarrow}0$ this reduces to the
Casimir energy of the non--rotating black hole
discussed in ref.\cite{Balasubramanian}.

Now we would like to show that the expression in
eqn.~(\ref{casimirtwo}) exactly matches the energy of the ${\cal
N}{=}4$ supersymmetric $U(N)$ Yang--Mills theory defined on a rotating
Einstein universe, which is the CFT on the boundary. The relation
between the parameters of the gravity theory in the bulk and those of
the CFT on the boundary is~\cite{Maldacena}
\bea
{ 1 \over G }={2 N^2\over \pi l^3}\ .
\label{dictionary}
\eea
Using this relation in eqn.~(\ref{casimirtwo}) one
gets
\be
{\cal E}={N^2  \over 48 l\Xi}(a^4/l^4+9 \Xi)\ .
\label{theenergy}
\ee

Since the the gravitational quasilocal stress tensor and the stress
tensor of the CFT on the boundary are dual to one another, one can use
the following expression for the energy of the field theory,
\bea
{\cal E}= \sum_{s=0,{1 \over 2},1} n^{s}\! \int_{\Sigma}d^3x \sqrt{\sigma}
\xi^{a}
<\!{{\widehat T}^{s}}_{ab}\!> u^{b}.
\eea
(Here $\xi^{a}$ and $u^{a}$ have the same meaning as before.)
This gives
\be
{\cal E}={4 \pi^2 \over \Xi}\!\!\!\sum_{s=0,{1 \over 2},1}\! \!n^s\!
\left[{\alpha^{s}\over 9}(3a^4/l^4-9\Xi)
+ {\beta^{s} \over 3}(a^4/l^4+9\Xi)\right]\ ,
\ee
where $n^s$ is the number of particles with spin $s$ in the ${\cal
N}{=}4$ super Yang--Mills theory on the boundary. The spin half
particle is the Weyl fermion. Substituting in the values for 
$(n^s,\alpha^s,\beta^s)$, 
we find that the resulting energy of the CFT 
exactly matches the prediction (\ref{theenergy}) 
for the Casimir energy from gravity.

\subsection{Conformal Anomaly}
If we expand the the trace of quasilocal stress tensor 
on the gravity side in
powers of
${1/r}$, the leading contribution is
\bea
{T}_{a}^{a}&=-&{a^2 l\over 8 \pi Gr^4}[ a^2/l^2
(3\cos^4{\theta}-2\cos^2{\theta})
-\cos{2\theta}] \nonumber\\ & &\hskip4cm  +O\left({1\over r^6}\right)\ .
\eea
Using the relation eqn.~(\ref{dictionary})\ between the gravitational
parameters and the gauge theory parameters and taking the large $r$
limit, one gets a prediction for the field theory quantity:
\be
{\widehat T}_{a}^{a}=-{N^2a^2 \over 4 \pi^2l^6}[a^2/l^2
(3\cos^4{\theta}-2\cos^2{\theta})
-\cos{2\theta}]\ .
\label{trace}
\ee
This precisely matches what one obtains by performing directly the
trace of the stress--tensor defined for the field theory in
eqn.(\ref{stresstwo}).

As a final check, we note that the general form of the conformal
anomaly in four dimensions (adapted to the present
case\cite{Henningson,Balasubramanian}) is given by
\bea
{\widehat T}_{a}^{a}=-{N^2\over 4\pi^2}
\left[-{1\over 8}{\cal R}^{\mu\nu}{\cal R}_{\mu\nu}+{1\over
24}{\cal R}^2\right]\ ,
\label{theanom}
\eea
where ${\cal R}_{ab}$ and ${\cal R}$ are the Ricci tensor and scalar
for the spacetime upon which the field theory resides.  Using the boundary
metric $\gamma_{ab}$, we find that this is precisely the result
obtained in eqn.(\ref{trace}). 

We find that the stress--tensor for the CFT on the rotating static
Einstein universe may be written in the following form:
\bea
{\widehat
T}^{ab}&=&A\,(4u^au^b+\gamma^{ab})+B\,v_{+}^av_{+}^b+
C\,v_{-}^av_{-}^b\nonumber\\
&+&D^+w_{+}^aw_{+}^b+D^-w_{-}^aw_{-}^b+E\,z^az^b+
{1\over4}\gamma^{ab}{\widehat
T}^d_d\ .
\label{stressform}
\eea
The unit time--like velocity $u^{a}$ and the accompanying null vectors are:
\bea
 u^{a}&=&(1,0,0,0)\ ,\nonumber\\
v_{\pm}^{a}&=&\left(1,0,0,\pm {1 \over l\cos\theta }\right)\ ,\quad
z^{a}=\left(1,-{2a\over l^2},0,{1 \over l\cos\theta }\right)\ ,\nonumber\\
w_{\pm}^{a}&=&\left(1,-{1\over l}\left(\pm1+{a\over l}\right),0,{1\over
l}\right)\ ,
\eea
while the coefficients are:
\bea
A&=&{1 \over 64\pi Gl}[a^4/l^4 (3\cos^4\theta -\cos 2\theta )
+8MG /l^2  \nonumber\\
&&\hskip5.5cm +2\Delta_\theta-1]\ ,\nonumber\\
B&=&{\cos\theta  \over 32\pi G l(1-\cos\theta)}
[a^4/l^4( 3\cos^4\theta-\cos^3\theta-\cos 2\theta) \nonumber\\
 &&\hskip5.5cm+2\Delta_\theta -1]\ ,\nonumber\\
C&=&-{a^2 \cos\theta \over
32\pi G l^3 }[a^2/l^2(3\cos^3\theta  -\cos\theta -\cos 2\theta ) \nonumber\\
 &&\hskip5.5cm +1-\cos\theta]\ ,\nonumber\\
D_\pm &=&-{a(1\pm a/l)\over 32\pi G l^2 (1-\cos\theta)}[a^3/l^3( \cos 2\theta
\cos\theta+\sin^2\theta )\nonumber\\
&&  \hskip4.5cm +a\cos\theta/l \pm\Delta_\theta]\ ,\nonumber\\
E&=&-{\cos\theta  \over
32\pi G l(1-\cos\theta )}\Xi\Delta_\theta\ .
\eea
Notice again, just as we saw in lower dimensions, that the coefficient, $A$, 
of the part of the tensor in standard form is the only one which
depends upon the thermal part of the theory ---represented by the black
hole mass--- while the other coefficients, referring to the null
vectors, are independent of $M$. It is also interesting to note that
when $a{\to}0$, while $C$ and $D^\pm$ vanish, the terms involving
$B$ and $E$ cancel each other.

The last term in the energy--momentum tensor (\ref{stressform})
appears to imply the presence of a conformal anomaly, which is
certainly interesting.  General considerations (see {\it e.g.,}
ref.\cite{Henningson} for a discussion in the present context) show
that in the presence of a conformal anomaly, the action has a
divergent piece 
\be 
I_{\rm div}\propto \log(r/l)\int d^4x\sqrt{-\gamma}\,
 {\widehat T}^a_a\ .  
\ee 
(See ref.\cite{counter}
for specific examples in this context.)  {}From reading
ref.\cite{robcliff}, however, one expects that the corresponding
logarithmic (UV) divergence ($\log(r/l)$) of the action should not be
present for a spacetime which can be written locally as a product. In
support of this, the action (\ref{theaction}) which we computed does
not have a logarithm (in contrast to the examples in
ref.\cite{counter}), as we have seen, presenting us with a
paradox. This is resolved upon realization that the {\it integrated}
trace actually does vanish, as a computation reveals\footnote{We are
grateful to R. C. Myers for a crucial discussion, and for pointing out
the vanishing.}. If this alone were true, it would provide a
counterexample\cite{paul,joe} to the often quoted folklore that, in a
non--trivial theory, scale invariance (preserved here) implies
conformal invariance, which is apparently broken here\footnote{We
thank P.~Townsend for pointing this out, and N.~Seiberg and
J.~Polchinski for discussions of this issue.}(see note added).

Unfortunately, this is not the case. A first clue that a manifestly
conformally invariant form for the stress tensor can be restored is
suggested by the fact\cite{Hawkingtwo} that the rotating Einstein
universe is conformal (up to an identification on the angular
coordinates) to the static Einstein universe after a change of
coordinates.  Furthermore, a closer examination of the expression for
the trace ${\widehat T}^a_a$ reveals that it is not only a total
derivative, but it is in fact proportional to $\Box {\cal R}$.  Since
a term proportional to $\Box {\cal R}$ results from a variation of
\begin{equation}
\Delta I_{\rm ct}={\kappa l^3 \over 8 \pi G}
\int_{\partial{\cal M}}{\cal R}^2\ ,
\end{equation} we can
therefore supplement the counterterms (\ref{theterms}) with such a
term, and define a new\footnote{In order to define ${\Theta}^{ab}$
with a suitably modified version of expression (\ref{stressone}), we
need to subtract an ``improvement term'', in the sense of
refs.\cite{improve,joe} which we will report on
elsewhere\cite{forth}.}  stress tensor ${\Theta}^{ab}$ which is
actually traceless for $\kappa{=}1/864$. Our stress tensor
${\Theta}^{ab}$ also has the same energy as
${\widehat T}^{ab}$(see note added).

This addition of a new counterterm ---which amounts to a new
regularisation scheme--- in order to ensure a manifestly conformally
invariant form for the stress tensor (which is not necessary when
there is no rotation in AdS$_5$), is an example of a fact familiar
from field theory: There is an ambiguity of the definition of the
trace anomaly (\ref{theanom}) in four dimensions, allowing a term
proportional to $\Box{\cal R}$ to be added, with an undetermined
coefficient corresponding to the choice of different schemes for
regularising the field theory\cite{Birrel}. Here, this corresponds to
different choices for boundary counterterms\cite{Balasubramanian}.  It
is interesting that a new counterterm is needed to restore manifest
conformal invariance after a change of variables {\it and} conformal
frame. This merits further investigation.

Overall, we see that the gravity computation (which is dual to a
computation in {\it strongly coupled} field theory) again matches the
results from the weakly coupled field theory computation. In the case
of pure AdS$_5$, the reason for the precise match, as explained in
ref.\cite{Balasubramanian}, is that pure AdS$_{5}$ receives no stringy
corrections because all tensors (in terms of which the stringy
corrections can be written) that can modify Einstein's field equations
vanish for that space.  This translates into no difference between the
results for strong and weak coupling field theory, because the stringy
corrections are correlated with corrections in field theory coupling,
as required by the duality\cite{Maldacena}.

That this is also true for the Kerr--AdS$_5$ case, in the limit of
zero black hole mass ---where we have shown full agreement with field
theory--- follows simply because $M{=}0$ Kerr--AdS$_5$ is just AdS$_5$
in non--standard coordinates\cite{Hawkingtwo}, parameterised by
$a$. It therefore shares many of the conformal properties with
AdS$_5$. As the stringy corrections can all be written in terms of the
Weyl tensor of the spacetime, if they vanish for AdS$_5$, they vanish
here also.

\section{Concluding Remarks}

The counterterm subtraction technique has allowed for the intrinsic
definition of the action (and the quantities which follow) for the
Kerr--AdS$_{n+1}$ spacetimes, for $n{=}2,3,4$.  This is in contrast to
the background subtraction technique, which required a reference
spacetime in order to compute a finite action.  In subtracting the
contribution from a reference spacetime to get a finite action, any
physics common to both spacetimes is lost. In this paper we have
computed examples of such quantities, the Casimir energies and
conformal anomaly for field theories residing on even dimensional
spacetimes. 

We have shown that the quantities thus computed (and indeed, the full
stress--energy--momentum tensor in four dimensions) are consistent
with a holographic interpretation of the physics of gravity in AdS in
terms of dual conformal field theory.  The Casimir energies which we
computed for the holographically dual two and four dimensional
conformal field theories (residing on a rotating cylinder, or the
rotating Einstein universe) exactly matched the contribution to the
energy computed for the Kerr--AdS spacetime which is not attributable
to the black hole. Also, we saw an interesting example of the change
in regularisation scheme needed to retain a manifestly conformally
invariant form in the four dimensional case. In all cases $(n{=}2,3,4)$ we
found that the stress tensor for the field theory on the rotating
spacetime can be neatly written in terms of a sum of a standard
traceless thermal form (involving the mass of the dual black hole) and
additional terms involving null vectors. Intriguingly, for $n{=}3$,
there are no such additional terms; its stress tensor keeps the simple
thermal form. This may be an important feature of the M2--brane
world--volume conformal field theory\cite{exoticone,exotictwo}. While
the computation for the two dimensional conformal field theory and the
related AdS$_3$ spacetime is essentially contained in
ref.\cite{Balasubramanian}, the results for higher dimensions are new.

We expect that these results also follow for Kerr--AdS$_{n+1}$, for
$n{+}1{>}5$. It would be especially interesting to see the form of the
tensor for Kerr--AdS$_7$.  However, we found that
the computational complexities encountered due to the off--diagonal
terms in the metric ---and the additional counterterms
needed~\cite{counter}--- were too severe to allow for an exploration
using the techniques which we have employed so far.

\section*{note added in proof} We noticed that while the new counterterm does
indeed 
restore the tracelessness of the the stress tensor, the new tensor thus
defined yields 
the same energy as the old but a different angular momentum. This appears to
means 
that the angular momentum of the Kerr black hole does break conformal
invariance
 of the theory, despite the fact that the rotating Einstein universe on the
boundary is
  conformal to Einstein universe. We intend to investigate this further and
report on 
  this matter in a future publication.

\section*{acknowledgements} We would like to thank Vijay Balasubramanian,
Roberto Emparan, Robert~C. Myers, Joe Polchinski, Harvey Reall, Nathan
Seiberg and Paul Townsend for comments and discussions, and especially
RE and RCM for suggestions and comments on a preliminary draft of this
paper.  CVJ would like to thank the Relativity group at D.A.M.T.P.,
Cambridge (UK) for hospitality during the early stages of this work.
This work was supported by an NSF Career grant, \#PHY--9733173. This
paper is report
\#'s UK/99--13, IASSNS--HEP--99/88, and DTP/99/71.

{}

\end{multicols}

\end{document}